\documentclass[prd,aps,twocolumn,nofootinbib,preprintnumbers,superscriptaddress,preprintnumbers,balancelastpage,longbibliography]{revtex4-2}
\usepackage[english]{babel}

\usepackage{amsmath,mathtools,physics,xfrac}
\usepackage{graphicx}
\usepackage{afterpage}
\usepackage{float}
\usepackage{subfigure}
\usepackage{rotating}
\usepackage{multirow}
\usepackage{tabularx}
\usepackage{booktabs}
\usepackage{fancyhdr}
\usepackage[utf8]{inputenc}
\usepackage{theorem}
\usepackage{moreverb}
\usepackage{euscript}
\usepackage{psfrag}
\usepackage{slashed}
\usepackage{mathtools}
\usepackage{makecell}
\usepackage{adjustbox}
\usepackage{dcolumn}
\usepackage{bm}
\usepackage[dvipsnames]{xcolor}
\usepackage{graphics}
\usepackage{hyperref}
\usepackage{chngcntr}
\usepackage{aas_macros}
\usepackage{tabularx}
\usepackage{color,xcolor}
\usepackage[normalem]{ulem}
\hypersetup{
     colorlinks   = true,
     citecolor    = Aquamarine,
     urlcolor     = Aquamarine,
     linkcolor    = Aquamarine
}

\definecolor{darkblue}{rgb}{0.0,0.0,0.75}
\definecolor{darkred}{rgb}{0.6,0.0,0}
\definecolor{darkgreen}{rgb}{0.0,0.6,0.}

\counterwithout{equation}{section}

\usepackage{tikz,xcolor,hyperref}

\definecolor{lime}{HTML}{A6CE39}
\DeclareRobustCommand{\orcidicon}{\hspace{-1mm}
	\begin{tikzpicture}
		\draw[lime, fill=lime] (0,0) 
		circle [radius=0.16] 
		node[white] {{\fontfamily{qag}\selectfont \tiny \,ID}};
		\draw[white, fill=white] (-0.0525,0.095) 
		circle [radius=0.007];
	\end{tikzpicture}
	\hspace{-3mm}
}

\foreach \x in {A, ..., Z}{\expandafter\xdef\csname orcid\x\endcsname{\noexpand\href{https://orcid.org/\csname orcidauthor\x\endcsname}
		{\noexpand\orcidicon}}
}


\begin{document}

\title{\boldmath Searching for Dark Matter with MeVCube}

\author{Akash Kumar Saha\orcidA{}}
\email{akashks@iisc.ac.in}

\affiliation{Centre for High Energy Physics, Indian Institute of Science, C.\,V.\,Raman Avenue, Bengaluru 560012, India}

	\date{\today}
	
	
	\begin{abstract}
CubeSat technology is an emerging alternative to large-scale space telescopes due to its short development time and cost-effectiveness. MeVCube is a proposed CubeSat mission to study the least explored MeV gamma-ray sky, also known as the `MeV gap'. Besides being sensitive to a plethora of astrophysical phenomena, MeVCube can also be important in the hunt for dark matter. If dark matter is made up of evaporating primordial black holes, then it can produce photons in the sensitivity range of MeVCube. Besides, particle dark matter can also decay or annihilate to produce final state gamma-ray photons. We perform the first comprehensive study of dark matter
discovery potential of a near-future MeVCube
CubeSat mission. In all cases, we find that MeVCube will have much better discovery reach compared to existing limits in the parameter space. This may be an important step towards discovering dark matter through its non-gravitational interactions.
  
	\end{abstract}
	
	\maketitle
	
\section{Introduction}
\label{Introduction}
The non-gravitational nature of dark matter (DM) remains one of the key fundamental questions of modern physics\,\cite{Bertone:2016nfn}. Despite having various gravitational evidences from different astrophysical and cosmological observables, so far all the search efforts to determine the nature of DM have returned null results\,\cite{Planck:2018vyg, Cirelli:2024ssz,Strigari:2012acq, Lisanti:2016jxe,Slatyer:2017sev, Lin:2019uvt}. Given $\sim$ 90 orders of magnitude spread in the possible DM mass range, one has to closely analyze all possible observables to arrive at any conclusion about the nature of DM.

One strategy is to look for Standard Model (SM) particles produced from DM, also known as indirect detection. For low-mass ($10^{15}$\,g - $10^{18}$\,g) Primordial Black Hole (PBH) DM, various Standard Model (SM) final states are produced due to Hawking evaporation\,\cite{Zeldovich:1967lct, 10.1093/mnras/152.1.75, 1974Natur.248...30H,Hawking:1975vcx,1975ApJ...201....1C}. These SM particles can then be detected by different observatories. Similarly, if DM has a particle nature, it can decay or annihilate to various SM final states and that will turn up as an `excess' in various astrophysical and cosmological observations.  Given the plethora of telescopes across all photon wavelengths, indirect search has resulted in stringent bounds for low-mass evaporating PBH DM and decaying/annihilating particle DM.

One of the least explored photon energy windows is $\sim$ 100 keV to 50 MeV. This is also known as the `MeV gap'\,\cite{Knodlseder:2016pey,Engel:2022bgx,Kierans:2022eid}. After the decommissioning of COMPTEL telescope (on board NASA’s
Compton Gamma-ray Observatory)\,\cite{1993ApJS...86..657S} in 2000, there has not been a dedicated telescope in this energy range and thus has remained poorly understood compared to other photon energies. There are many proposed telescopes like COSI\,\cite{Tomsick:2019wvo}, AMEGO-X\,\cite{AMEGO:2019gny,Fleischhack:2021mhc}, e-ASTROGAM\,\cite{e-ASTROGAM:2017pxr}, GECCO\,\cite{Orlando:2021get}, MAST\,\cite{Dzhatdoev:2019kay}, GRAMS\,\cite{Aramaki:2019bpi}, PANGU\,\cite{Wu:2014tya} and many others that can close this gap. Recently COSI has been selected for launch in 2027\,\cite{cosi}. This will mark the beginning of a new era for MeV astronomy and astrophysics.
\begin{figure}
	\begin{center}
	\includegraphics[height=7.8cm]{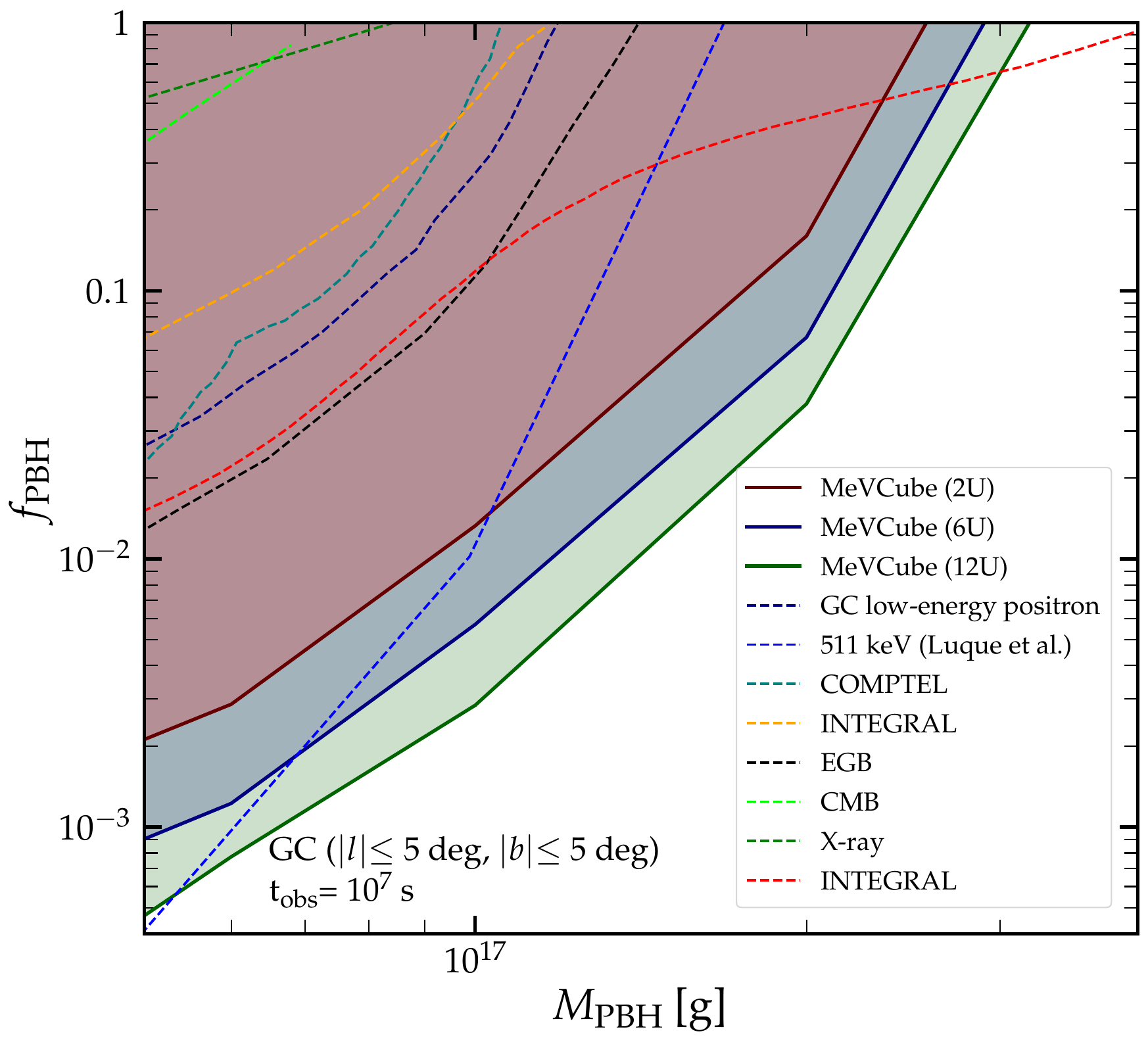}
	\caption{Sensitivities on the fraction of DM made up of PBH DM ($f_{\rm PBH}$) as a function of PBH mass ($M_{\rm PBH}$) with observations by MeVCube of sizes 2U (brown solid line), 6U (blue solid line), and 12U (green solid line).  Previous limits are shown by dashed curves. These include galactic 511 keV photon measurement by INTEGRAL (navy \cite{Laha:2019ssq} and blue \cite{DelaTorreLuque:2024qms}), measurements of the diffuse Galactic gamma-ray flux by COMPTEL (teal)\,\cite{PhysRevLett.126.171101} and INTEGRAL (orange\,\cite{Laha:2020ivk} and red\,\cite{Berteaud:2022tws}), extra-Galactic gamma-ray measurement  (black) \cite{PhysRevD.101.023010,Chen:2021ngo}, CMB measurements by PLANCK (lime) \cite{Clark:2016nst}, and Galactic diffuse X-ray emission measurement by XMM-Newton (green) \cite{DelaTorreLuque:2024qms}.}
	\label{fig:-PBH} 
     \end{center}
\end{figure}
\begin{figure*}[!htbp]
	\begin{center}
		\includegraphics[height=7.8cm]{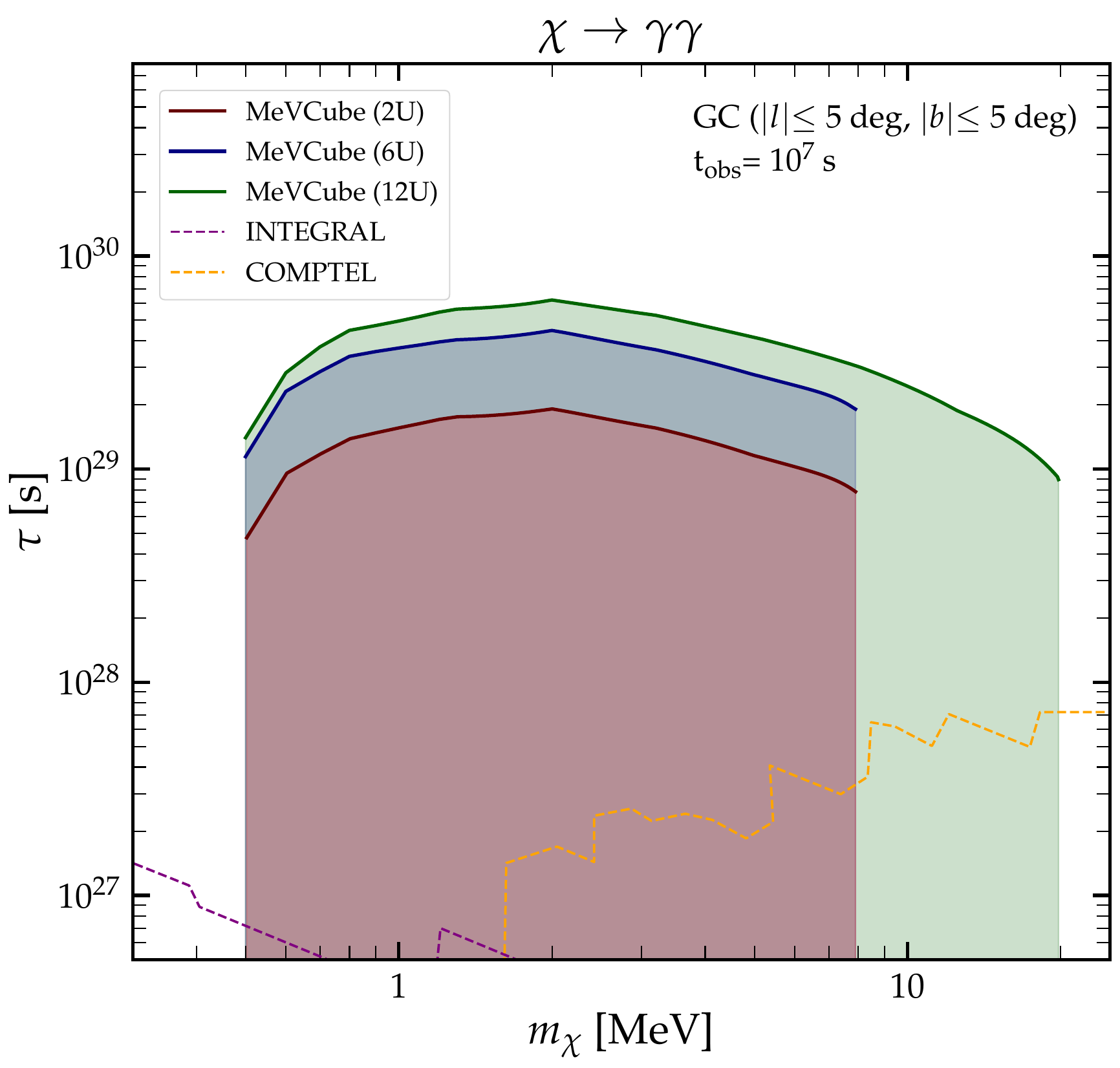}
		\includegraphics[height=7.8cm]{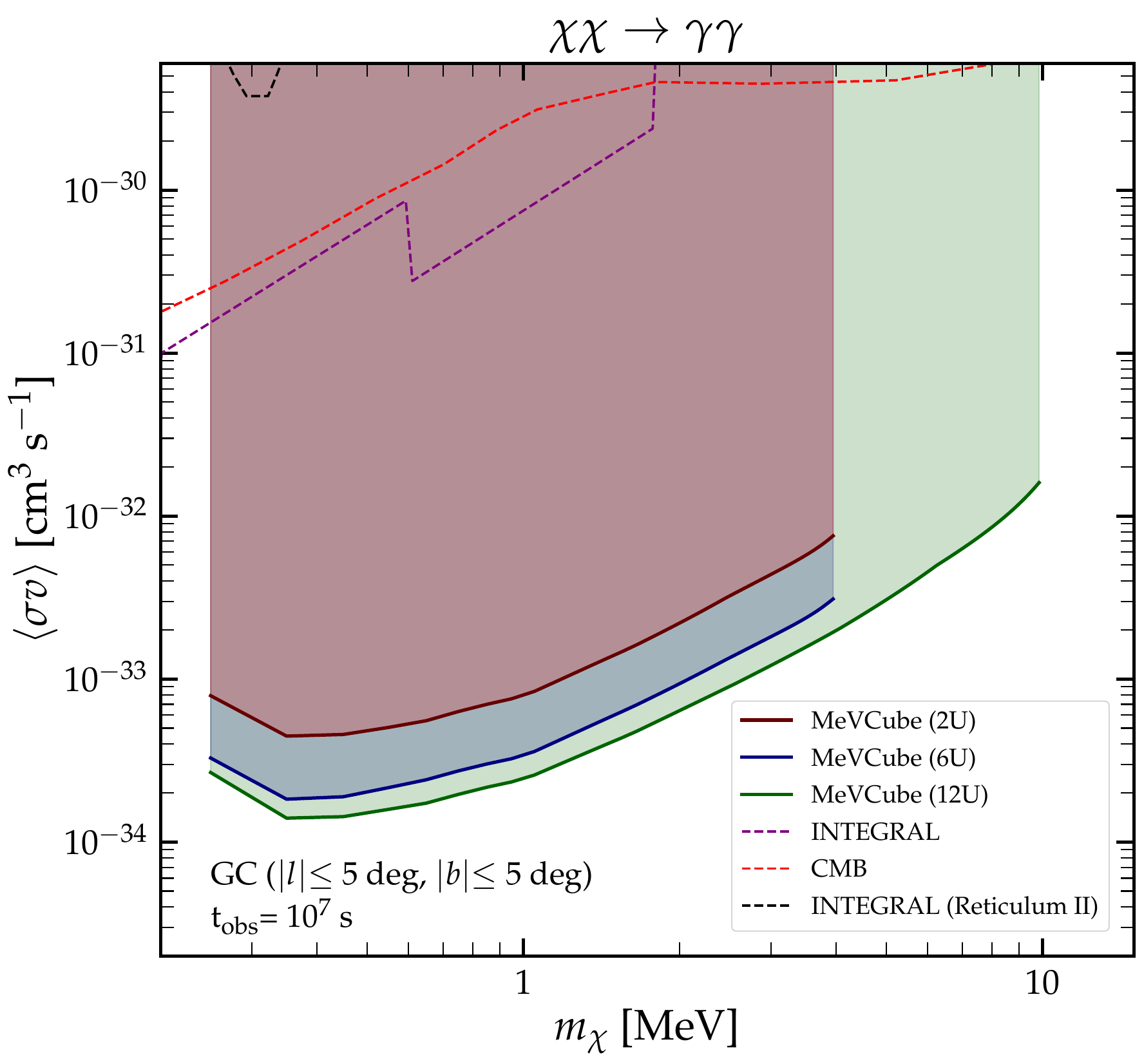}~~\\	
	
		\label{fig: -decay and annihilation}
	\end{center}	
\end{figure*}

\vspace{-0.001cm} 

\begin{figure*}[!htbp]
	\begin{center}
		\includegraphics[height=7.8cm]{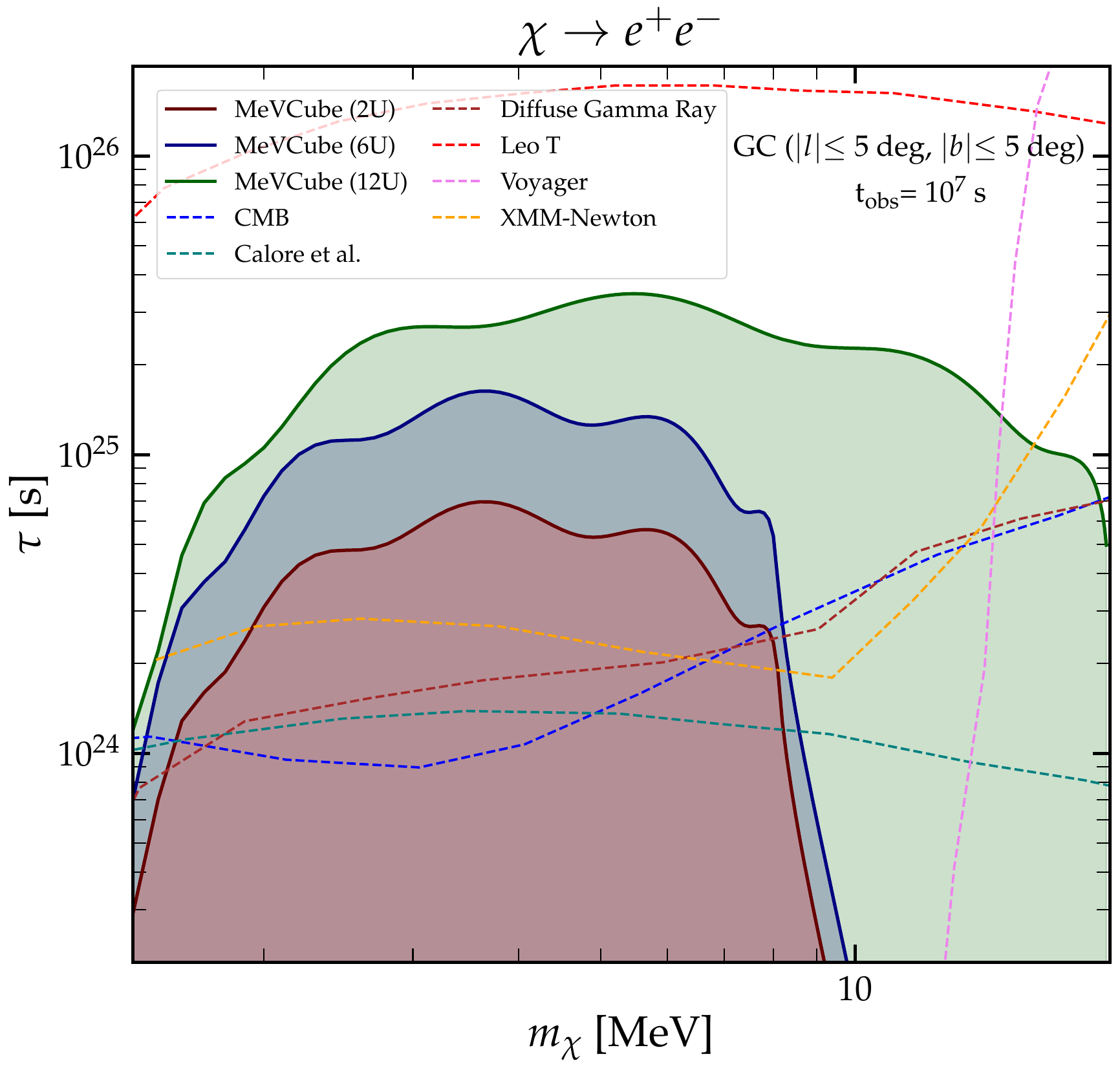}
		\includegraphics[height=7.8cm]{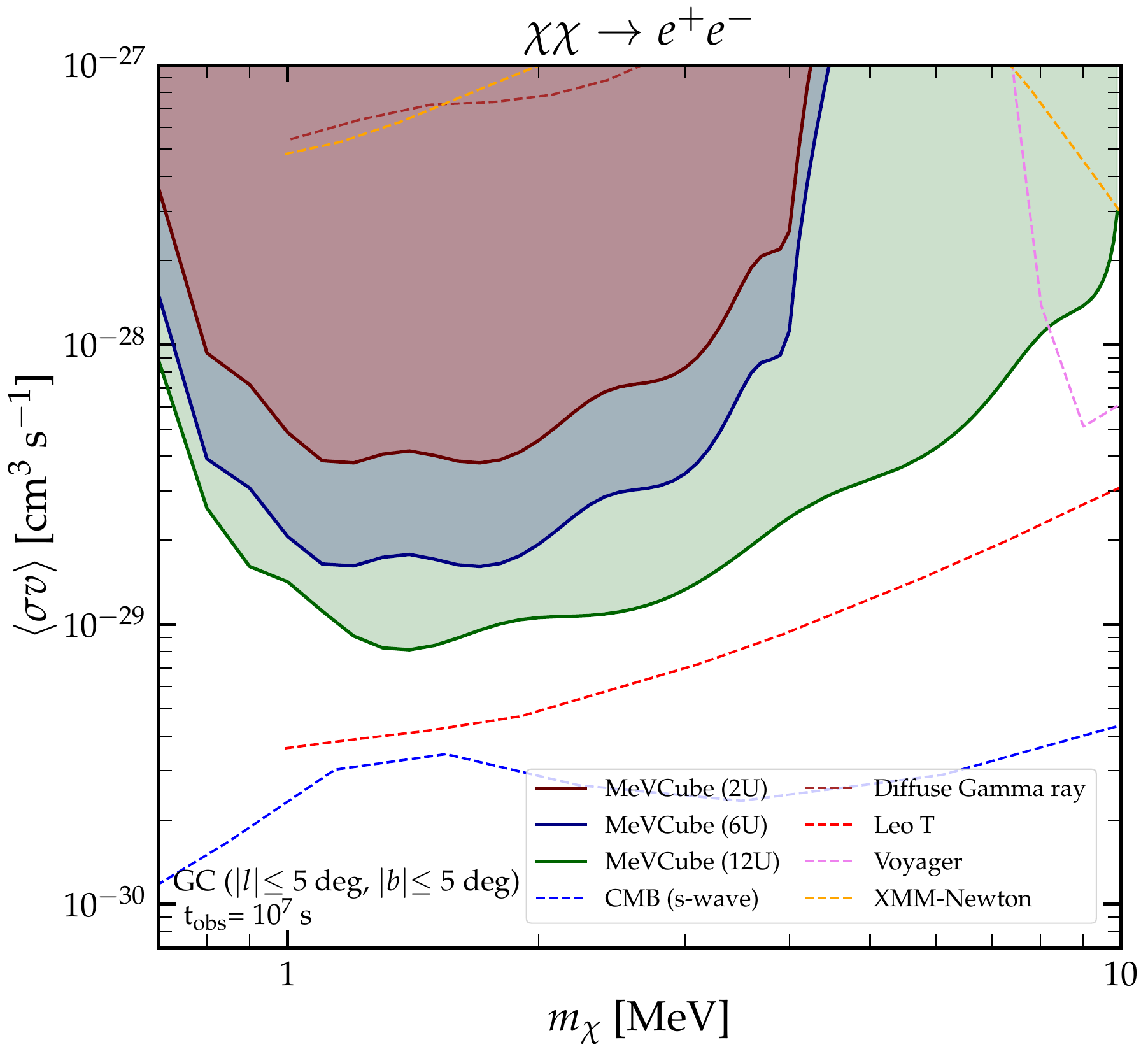}~~\\	
		\caption{(\textbf{Top row}) sensitivities on the lifetime (left) and velocity averaged cross-section (right) for particle DM with observations by MeVCube of sizes 2U (brown solid line), 6U (blue solid line), and 12U (green solid line). The existing limits include observations of diffuse galactic emission from INTEGRAL\,\cite{Laha:2020ivk,Siegert:2021upf} (purple dashed line) and COMPTEL\,\cite{Essig:2013goa} (orange dashed line), CMB anisotropy measurements\,\cite{Slatyer:2015jla} (red dashed line), and INTEGRAL observation of Reticulum II dwarf galaxy\,\cite{Siegert:2021upf} (black dashed line). (\textbf{Bottom row}) sensitivities on the lifetime (left) and velocity averaged cross-section (right) for particle DM (photon production via FSR) with observations by MeVCube of sizes 2U (brown solid line), 6U (blue solid line), and 12U (green solid line). For DM decay, previous limits include limits from CMB anisotropy measurement\,\cite{Liu:2016cnk} (blue dashed line), INTEGRAL/SPI diffuse gamma ray limits\,\cite{Calore:2022pks} (blue dashed line), combined limits from soft gamma-ray and X-ray data from\,\cite{Essig:2013goa} (brown dashed line), gas heating limits from LeoT\,\cite{Wadekar:2021qae} (red dashed line), limits utilizing Voyager 1 data\,\cite{Boudaud:2016mos} (violet dashed line), limits from diffuse X-ray emission seen by XMM-Newton\,\cite{Cirelli:2023tnx} (orange dashed line). For DM annihilation, the CMB limit is taken from Ref.\,\cite{Slatyer:2015jla} (blue dashed line). }
		\label{fig: -FSR decay and annihilation}
	\end{center}	
\end{figure*}
One of the problems faced by any upcoming telescope collaboration is the long planning duration and financial constraints. As a result, only a few satellites actually get launched in the end to achieve their corresponding science goals. In recent years, to circumvent these issues, the CubeSat technology has become a popular alternative. CubeSats are small-scale satellites where a 1U CubeSat is $10 \,{\rm cm}\times10\,{\rm cm}\times11.35\,{\rm cm}$ and weights around 1 kg (see `CubeSat Design Specification' (CDS) in\,\cite{cubesat}). Many such units can be combined to make a bigger CubeSat. Despite the small size, these CubeSats can compete with some of the large scale telescopes thanks to various technological advancements. Due to smaller size, development time and cost are also minimal. Besides, due to the smaller weight, the launching cost is also minimal compared to big scale telescopes\footnote{\href{https://www.spacex.com/rideshare/}{https://www.spacex.com/rideshare/}\\
\href{https://www.rocketlabusa.com/launch/electron/}{https://www.rocketlabusa.com/launch/electron/}
\\\href{https://www.isro.gov.in/StudentSatellite.html}{https://www.isro.gov.in/StudentSatellite.html}}. Thus, CubeSats offer a unique opportunity of studying astrophysical phenomena in a cost-effective way. 

Recently a proposal has been made to fill the MeV gap using a CubeSat telescope, called MeVCube\,\cite{Lucchetta:2022nrm,Lucchetta:2022zrd}. With two layers of CdZnTe semiconductor devices, the sensitivity of the telescope is shown to be comparable or in some cases better than that achieved by COMPTEL. This opens up the possibility of building and launching these small MeVCube telescopes before the big scale proposals. This can pave the way for the next generation full-scale telescopes.

Given the effectiveness of MeVCube in unraveling the MeV sky, we ask the question: can MeVCube search for DM? In this work, for the first time, we investigate the reach of MeVCube as a DM detector. Previously the New Physics capabilities of various proposed MeV satellites have been performed\,\cite{Boddy:2015efa,Ray:2021mxu,Coogan:2020tuf,Coogan:2021sjs,Caputo:2022dkz,Coogan:2021rez,Ghosh:2021gfa,Tseng:2022jta,Carenza:2022som,Xie:2023cwi,Berlin:2023qco,Calza:2023rjt,Kasuya:2024ldq,Cui:2024uwk,ODonnell:2024aaw,Dent:2024yje,Xie:2024eug,Agashe:2024dkq,Compagnin:2022elr,Berlin:2023qco,JUNO:2023vyz,Boddy:2024cyu,Manzari:2024jns,Alpine:2024kej,Buckley:2024ldr}. In each case, the upcoming telescopes are shown to be excellent tools for DM search. This work explores the power of CubeSat technology in the search for DM.

This paper is structured as follows. Section \ref{DM models} gives an overview of the three different DM models from which MeV photons can be produced, namely, PBH DM and decaying, annihilating particle DM. We then discuss the proposed MeVCube satellite and how it will be able to fill the MeV gap in section \ref{MevCube}. In section \ref{results} we show our results and finally discuss  future scopes in section \ref{scope}.

\section{MeV photons from different DM models}
 \label{DM models}
Many well-motivated models in the literature have explored the possibility of having MeV scale DM and production of photons from them\,\cite{Boehm:2003hm, Pospelov:2007mp, Hochberg:2014dra, Boyarsky:2018tvu, Evans:2019jcs,  Dasgupta:2021ies, Compagnin:2022elr, Chu:2022xuh,Linden:2024fby,Nguyen:2024kwy,Balan:2024cmq}. Here we explore such possibilities in a model-independent way. We will consider two classes of DM models that can produce photons that MeVCube will be sensitive to: PBH DM and particle DM. 

A DM with mass $M_\chi$ can decay or annihilate to produce a photon flux given by,
\begin{eqnarray}
    \phi_{\rm DM}(E_\gamma)=\frac{1}{2^{\alpha-1}}\frac{r_\odot}{4 \pi}\frac{\rho_\odot}{M_\chi}\left(\frac{d^2N}{dE_\gamma dt}\right)\frac{J}{\Delta\Omega},
    \label{fluxperenergy}
\end{eqnarray}
in units of MeV$^{-1}$cm$^{-2}$s$^{-1}$sr$^{-1}$, where $\alpha=1(2)$ for DM decay (annihilation), $\rho_\odot$ = 0.4 GeV cm$^{-3}$ is the DM density in solar neighborhood, $r_\odot$ = 8.3 kpc is the distance to the Milky Way (MW) centre from the Solar position, $\left(\frac{d^2N}{dE_\gamma dt}\right)$ is the spectra of photons produced per decay or annihilation per unit time, $\Delta\Omega$ is the angular size of the region we are looking at, and $J$ is the so called D-factor (for DM decay) or J-factor (for DM annihilation). This factor encapsulates the astrophysical distribution of DM and is given by,
\begin{eqnarray}
   J = \int \frac{ds}{r_\odot}\left(\frac{\rho}{\rho_\odot}\right)^\alpha d\Omega\,,
\end{eqnarray}
where $\rho$ is the DM density profile in the MW halo and $s$ is the line of sight distance. We note that the above expressions also holds for PBH DM candidate, as a PBH can be thought of as a `decaying' DM.

The density profile of DM in our MW halo can be parametrized as\,\cite{Bringmann:2012ez},
\begin{eqnarray}
\rho_\chi^{a b c} (r) \,=\, \rho_\odot \, \left(\dfrac{r}{r_\odot} \right)^{- c} \, \left(\dfrac{1 \,+\, (r_\odot/r_s)^a}{1 \,+\, (r/r_s)^a} \right)^{(b - c)/a}\,,
\label{NFW}
\end{eqnarray}
where $r$ is the distance from the Galactic Center (GC), and $r_s$ is the scale radius. Here, the distance 
$r$ is related to the line of sight distance $s$ via,
\begin{eqnarray}
    r(s,\,b,\,\ell) \,=\, \sqrt{s^2 \,+\, r_\odot^2 \,-\, 2 \, s \, r_\odot \, {\rm cos}\, b \, {\rm cos}\,\ell}\,\,,
\end{eqnarray}
where $l$ and $b$ are the galactic longitude and latitude of the region of interest, respectively. The differential angular size then becomes, $d\Omega$ = $d$\,(sin\,$b$)\,$d\ell$.
For this work, we use the widely considered NFW profile as benchmark DM profile, with $(a,b,c)=(1,3,1)$, and $r_s$ = 20 kpc in Eq.\,(\ref{NFW}). The effect of other DM density profiles on our results is shown in Appendix B. Next, we look at the spectra of photons from DM models considered here.

\subsection{PBH DM evaporation}
 \label{PBH review}
PBHs can be formed in the very early universe due to various different mechanisms\,\cite{Shibata:1999zs, Harada:2013epa, Musco:2023dak, Harada:2016mhb, Niemeyer:1999ak, Musco:2018rwt, Carr:1974nx,Yoo:2020lmg,Clesse:2015wea, Jedamzik:1996mr,Bhattacharya:2019bvk,Bhaumik:2019tvl, Bhaumik:2020dor,Escriva:2019phb,Escriva:2022bwe,Escriva:2024lmm}. After formation, PBHs will Hawking  evaporate\,\cite{Hawking:1975vcx} depending on their mass, electric charge, and angular momentum. For an uncharged, non-spinning black hole, the temperature is given by,
\begin{eqnarray}
T_{\rm PBH}=\frac{1}{8\pi G M_{\rm PBH}}=1.06\left(\frac{10^{13} \,\text{g}}{M_{\rm PBH}}\right) \text{GeV},
\label{eqn:1}
\end{eqnarray}
where $M_{\rm PBH}$ is the mass of the PBH and $G$ is the gravitational constant. 

The number of emitted particles per unit energy per unit time from an evaporating, uncharged, non-spinning black hole is given by\,\,\cite{Page:1976df,Page:1976ki,MacGibbon:1990zk,Arbey:2019mbc}
\begin{eqnarray}
\frac{d^2N_i}{dE dt}=\frac{1}{2 \pi}\sum_{dof}\frac{\Gamma_i(E,M_{\rm PBH},\mu)}{e^{E/T}- (-1)^{2s}},
\end{eqnarray} 
where $\Gamma_i$ are the greybody factors that encode the probability of an emitted particle to surpass the gravitational potential of the black hole, $E$, $\mu$, $s$ are the total energy, rest mass and spin of the emitted particle, respectively. The symbol $dof$ specifies the degrees of freedom associated with the emitted particle. For only photon final state the above expression simplifies,
\begin{eqnarray}
 \frac{d^2N}{dE_\gamma dt}=\frac{1}{2 \pi}\frac{\Gamma(E_\gamma,M_{\rm PBH})}{e^{E_\gamma/T}- 1},   
\end{eqnarray}
where $E_\gamma$ is the energy of the emitted photons and $\Gamma$ contains an additional factor of 2 corresponding to two degrees freedom for photon.

We evaluate the spectra of primary and secondary photons from low-mass evaporating PBH using publicly available package $\tt{BlackHawk}$\,\cite{Arbey:2019mbc,Arbey:2021mbl}. We have extensively cross-checked the output of $\tt{BlackHawk}$ with the semi-analytical forms of
greybody factor provided in Ref.\,\cite{Page:1976df}. To evaluate secondary photon flux at lower energies, $\tt{BlackHawk}$ utilizes $\tt{Hazma}$\,\cite{Coogan:2019qpu}. We use the sum of both primary and secondary photon as the signal flux from PBH DM.

 \subsection{Particle DM Decay and Annihilation}
Now we focus on the case of decaying and annihilating particle DM and the resulting photon spectra.

For a particle DM of mass $m_\chi$ directly decaying to two photons, the resulting photon spectra is given by,
\begin{eqnarray}
    \frac{d^2N}{dE_\gamma dt}=\frac{2}{\tau}\delta\left(E_\gamma-\frac{m_\chi}{2}\right)\,\,,
    \label{decay}
\end{eqnarray}
where $\tau$ is the DM lifetime, and E$_\gamma$ is the energy of the emitted photon. 

In case of annihilating DM, the photon spectra becomes,
\begin{eqnarray}
     \frac{d^2N}{dE_\gamma dt}=2 \langle\sigma v\rangle \left(\frac{\rho_\odot}{m_\chi}\right) \delta\left(E_\gamma-m_\chi
     \right)\,\,,
     \label{annihilation}
\end{eqnarray}
where $\langle\sigma v\rangle$ is the velocity averaged DM annihilation cross-section. The factor of 2 in Eqs.\,(\ref{decay},\,\ref{annihilation}) arise because per DM decay and annihilation we have production of two photons.

 Besides direct decay and annihilation to photons, DM particles can also decay or annihilate to electron-positron pairs, which in turn can produce photons via Final State Radiation (FSR). The flux of the resulting photons is given by\,\cite{Essig:2013goa, Cirelli:2020bpc},
 \begin{eqnarray}
     \frac{d^2N}{dE_\gamma dt}= \frac{2\,\mathcal{C}_\alpha\,\alpha_e}{\pi\beta(3-\beta^2)\alpha m_\chi}\left(A\,{\rm ln}\left(\frac{1+R(\nu)}{1-R(\nu)}\right) - 2 B R(\nu)\right)\, \nonumber
 \end{eqnarray}
with,
\begin{eqnarray}
    A=\frac{(1+\beta^2)(3-\beta^2)}{\nu}-2(3-\beta^2) + 2\nu\,\,,
\end{eqnarray}
\begin{eqnarray}
    B=\frac{(1-\nu)(3-\beta^2)}{\nu} +\nu\,\,,
\end{eqnarray}
where,
\begin{eqnarray}
    \mu=\frac{m_e}{\alpha m_\chi},\,\, \beta=\sqrt{1-4\mu^2},\,\,\nu=\frac{2E_\gamma}{\alpha m_\chi}\,\,,
\end{eqnarray}
and,
\begin{eqnarray}
    R(\nu)=\sqrt{1-4\left(\frac{\mu^2}{1-\nu}\right)}\,\,,
\end{eqnarray}
where $\alpha_e$ is the fine structure constant and again $\alpha =1 (2)$ for DM decay (annihilation). Accordingly, the factors $\mathcal{C}_1=\frac{1}{\tau}$ (for DM decay), and $\mathcal{C}_2=\langle\sigma v\rangle\left(\frac{\rho_\odot}{m_\chi}\right)$ (for DM annihilation). 

For DM annihilation,  we assume that DM is made up of self-conjugated particles. If DM is not its own anti-particle then an additional factor of $1/2$ has to be included in Eq.\,(\ref{fluxperenergy}). 

For all particle DM decay and annihilation channels we assume
100\% branching ratio for simplicity.

We note that for the cases of low-mass evaporating PBH DM and particle DM decaying or annihilating to photons via FSR, the resulting signal is a continuum spectra of photons. For particle DM directly decaying or annihilating to photons, the signal is a line feature. Besides, for some particular DM models we can have $\chi\rightarrow\phi\gamma$,\,\,$\chi\chi\rightarrow\phi\gamma$,\,\,$\chi\chi\rightarrow\phi\phi\rightarrow4\gamma$ for some state $\phi$\,\cite{Goodman:2010qn,Rajaraman:2012db,Bergstrom:2012vd,Ibarra:2012dw}. The distinct photon spectral features for different DM models result in different sensitivities for DM search with  existing and upcoming telescopes.

 \section{CubeSat for MeV gap: MeVCube}
\label{MevCube}
MeVCube\,\cite{Lucchetta:2022nrm,Lucchetta:2022zrd} is a proposed `CubeSat' standard MeV telescope for exploring the `MeV gap'. The detector will consist of two layers of pixelated Cadmium-Zinc-Telluride (CdZnTe) detector with low-power read-out electronics. The distance between the two layers is 6 cm. Here the first layer acts as a scattering layer, whereas the second layer acts as an absorption layer. Gamma-rays in $\sim$ MeV energies primarily interact with matter via Compton scattering. Photons incident on the first layer of the MeVCube scatter, lose energy, and are absorbed in the second layer. By measuring the energies of the scattered electron in the first layer and the energy of the absorbed photon in the second layer, along with the interaction site, one can reconstruct the direction and energy of the initial photon\,\cite{Kierans:2022eid}. COMPTEL also used a `double scatter' scheme where the first layer was a low-Z scintillator (NE 213A) and the second layer was a high-Z scintillator (NaI(Tl))\,\cite{1993ApJS...86..657S}. We note that COSI will use 6 double-sided strip germanium detectors with independent readouts\,\cite{Tomsick:2019wvo}.

The final size of the MeVCube is yet to be decided. Following Ref.\,\cite{Lucchetta:2022nrm}, we look at three different units of MeVCube, 2U, 6U, and 12U. The 2U and 6U versions of MeVCube will be sensitive to photons in the energy range $\sim$ 200 keV to 4 MeV (for a 12U CubeSat it extends upto $\sim$ 10 MeV). MeVCube will have a wide field-of-view ($\sim 2$ sr), and can achieve an angular resolution of 1.5$^{\circ}$ at 1 MeV.

Despite having smaller size compared to COMPTEL, MeVCube will achieve comparable effective area, mainly due to better event reconstruction capability of initial photon direction and energy. CdZnTe has a high atomic
number and a wide energy gap ($\sim$ 1.52 eV) with low leakage current\,\cite{2022NIMPA103766922L}. As a result, CdZnTe has high spectral and imaging
capabilities at room temperature\,\cite{SCHLESINGER2001103}. Besides MeVCube utilizes VATA450.3 ASICs readouts that can work
with much less power and is ideal for CubeSat missions\,\cite{Lucchetta:2022nrm}. Pixelated CdZnTe and individual read-outs
are also ideal for reconstruction of incident photon directions as opposed to having one single target\,\cite{Pixelated}. These are some of the technological advancements that have been crucial for MeVCube sensitivity. The effective area and energy resolution for MeVCube is obtained via detailed simulations calibrated to the in-lab measurements of CdZnTe prototype, and is provided in Ref.\,\cite{Lucchetta:2022nrm}.

MeVCube stands unique in the fleet of all the proposed MeV satellites. Due to the small size, the development cost and time is much less compared to any of the big scale telescopes. Given the smaller weight, the launch cost of MeVCube is also expected to be less than any other proposed MeV satellites. 

 The primary science goals of MeVCube will be to search for Gamma-ray Bursts (GRBs), spectral lines from galactic nucleosynthesis , electromagnetic counterparts of gravitational wave (GW) and many other astrophysical phenomena that can produce MeV photons. Here we use, for the first time, the preliminary specifications provided in Ref.\,\cite{Lucchetta:2022nrm}, to check MeVCube's sensitivities for low-mass evaporating PBH DM and decaying, annihilating particle DM. Given the rise of CubeSat technology in traditional space science, the understanding of DM physics capabilities of CubeSats will be of great importance in the near future. 
 
 \section{Results}
 \label{results}
For this work we focus on MW GC region defined by $|l|\leq5^\circ$, $|b|\leq5^\circ$. We parametrize the galactic and extragalactic photon backgrounds following Refs.\, \cite{Bartels:2017dpb,Beacom:2005qv,Ray:2021mxu}, where the galactic background is given by,
 \begin{eqnarray}
     \phi_{\rm g}^{\rm bkg}(E_\gamma)= A_{\rm g}^{\rm bkg}\left(\frac{E_\gamma}{1\,\rm MeV}\right)^{-\alpha^{\rm g}}{\rm exp}\left(-\left(\frac{E_\gamma}{E_c}\right)^{\bar{\gamma}}\right)
     \label{galbkg}
 \end{eqnarray}
 and the extragalactic background component is,
 \begin{eqnarray}
     \phi_{\rm eg}^{\rm bkg}(E_\gamma)=A_{\rm eg}^{\rm bkg}\left(\frac{E_\gamma}{1\,\rm MeV}\right)^{-\alpha^{\rm eg}}\,\,,
      \label{exgalbkg}
 \end{eqnarray}
both of which are in units of MeV$^{-1}$cm$^{-2}$s$^{-1}$sr$^{-1}$, and where the fiducial background parameter values are, $A_{\rm g}^{\rm bkg}= 0.013$ MeV$^{-1}$cm$^{-2}$s$^{-1}$sr$^{-1}$, $\alpha^{\rm g}= 1.8$, $\bar{\gamma}=2$, $E_c=20$ MeV\,\footnote{We note there is a typo in Eq.\,(4.1) of Ref.\,\cite{Bartels:2017dpb}, where the cut-off energy is given as $E_c=2$ MeV. We briefly discuss this in Appendix C.} , $A_{\rm eg}^{\rm bkg}=0.004135$ MeV$^{-1}$cm$^{-2}$s$^{-1}$sr$^{-1}$, $\alpha^{\rm eg}=2.8956$.
For the galactic background, best-fit parameter values are obtained by comparing with the COMPTEL data in the region $|l| < 30^{\circ}, |b| < 5^{\circ}$\,\cite{2005A&A...444..495S,Strong:1998ck,Beacom:2005qv,Bartels:2017dpb}. For extragalactic background, the fiducial values are obtained by fitting the measured
cosmic X-ray background spectrum in the energy range 150 keV to 5 MeV\,\cite{Ray:2021mxu,Ballesteros:2019exr}. We note that Refs.\,\cite{Ray:2021mxu, ODonnell:2024aaw} have used the correct value of $E_c$ (i.e., 20 MeV) in their works. We elaborate on this in Appendix C.

The DM flux can be calculated using the spectra evaluated in Sec.\,\ref{DM models} for each of the DM models. The spectrum of photons measured by MeVCube will be modified due to the finite energy resolution of the instrument\,\cite{Bringmann:2008kj},
\begin{eqnarray}
   \left(\frac{d^2N}{dE_\gamma dt}\right)^\prime(E_\gamma) = \int dE_\gamma^\prime \mathcal{R}_\epsilon(E_\gamma-E_\gamma^\prime)\left(\frac{d^2N}{dE_\gamma dt}\right)(E_\gamma^\prime)\,\,,\nonumber
\end{eqnarray}
where,
\begin{eqnarray}
    \mathcal{R}_\epsilon(E_\gamma-E_\gamma^\prime) = \frac{1}{\sqrt{2 \pi(\epsilon E_\gamma)^2}}{\rm exp}\left(\frac{-(E_\gamma-E_\gamma^\prime)^2}{2 (\epsilon E_\gamma)^2}\right)
\end{eqnarray}
In the above equation $\mathcal{R}_\epsilon(E_\gamma-E_\gamma^\prime)$ is the energy resolution of the telescope, specifying the probability that a photon of true energy $E_\gamma^\prime$ will be detected as a photon of energy $E_\gamma$. We use the full width at half maximum (FWHM) energy resolution provided for MeVCube in Ref.\,\cite{Lucchetta:2022nrm} (Fig.\,4). We note that the $\epsilon$ parameter defined above relates to the FWHM (\%) resolution given in Ref.\,\cite{Lucchetta:2022nrm} by,
\begin{eqnarray}
    \epsilon = \frac{\rm FWHM}{100}\times\frac{1}{2.355}\,.
\end{eqnarray}

The telescope energy resolution response is important for line searches and the effect is small for continuum signals. With the resolution taken into account, we then use Eq.\,(\ref{fluxperenergy}) to calculate the total flux of photons emitted for a particular DM model.

The total flux is the sum of DM origin photon flux and the background photon flux, 
 \begin{eqnarray}
     \Phi(\theta)=\phi_{\rm DM}(\theta_{\chi}) + \phi^{\rm bkg} (\theta_{\rm bkg})\,\,,
 \end{eqnarray}
 where $\theta_{\chi}$ and $\theta_{\rm bkg}$ are the DM model and astrophysical background parameters, respectively. For PBH DM, decaying DM and annihilating DM the model parameters are, $\theta_{\chi}$= ($f_{\rm PBH}$, $1/\tau$, $\langle\sigma v\rangle$), respectively, for  a particular DM mass and decay/annihilation channel. The background flux includes both galactic and extra-galactic components,
 \begin{eqnarray}
     \phi^{\rm bkg} (\theta_{\rm bkg})= \phi_{\rm g}^{\rm bkg}(\theta_{\rm bkg}) +  \phi_{\rm eg}^{\rm bkg}(\theta_{\rm bkg})\,\,,
 \end{eqnarray}
where $\theta_{\rm bkg}= (A_{\rm g}^{\rm bkg}, \alpha^{\rm g},\bar{\gamma}, E_c,  A_{\rm eg}^{\rm bkg}, \alpha^{\rm eg}) $ denotes the background model parameters.

For the DM signal, here we only consider galactic DM contributions. In principle, extragalactic DM signal will also be present in the total photon flux detected. For our region of interest ($|l|\leq5^\circ$, $|b|\leq5^\circ$) the latter contribution is subdominant.

We apply Fisher forecasting method\,\cite{Edwards:2017mnf} and marginalize over all the background parameters to obtain limits on DM model parameters. The Fisher matrix elements are given by,
\begin{eqnarray}
    \mathcal{F}_{ij}=\int dE_\gamma\,d\Omega\,t_{\rm obs}\,A_{\rm eff}(E_\gamma)\left[\frac{1}{\Phi}\left(\frac{\partial\Phi}{\partial \theta_i}\right)\left(\frac{\partial\Phi}{\partial \theta_j}\right)\right]_{\theta=\theta_{\rm fid}}\,\,
\end{eqnarray}
where $t_{\rm obs}$ is observation time, $A_{\rm eff}(E_\gamma)$ is the effective area of the telescope, and $\theta_{\rm fid}$ is the fiducial value for a particular parameter $\theta$. For DM signal, we set $(\theta_\chi)_{\rm fid}=0$. We choose $t_{\rm obs}= 10^7$ s as our benchmark observation time. The variation in our limits due to different values of $t_{\rm obs}$ is shown in Appendix B. The effective area, $A_{\rm eff}(E_\gamma)$ is taken from Fig.\,5 of Ref.\,\cite{Lucchetta:2022nrm}. 

In our case, $\mathcal{F}$ is a 7$ \times 7$ symmetric matrix with 1 model parameter $(f_{\rm PBH}$ or $(1/\tau)$  or $\langle\sigma v\rangle)$ and 6 background parameters ($A_{\rm g}^{\rm bkg}, \alpha^{\rm g},\bar{\gamma}, E_c,  A_{\rm eg}^{\rm bkg}, \alpha^{\rm eg} $). Assuming that MeVCube during its observation time does not observe any DM signal, we can put 95\% confidence level upper limit on the DM model parameter $\theta_{\chi}$ as\,\cite{Edwards:2017mnf},
\begin{eqnarray}
    \theta_{\chi}^{\rm UL} = 1.645 \sqrt{(\mathcal{F}^{-1})_{11}}\,\,,
\end{eqnarray}
where $\theta_{\chi}=(f_{\rm PBH}, 1/\tau, \langle\sigma v\rangle)$ for PBH evaporation, decaying DM, and annihilating DM, respectively. For low-mass evaporating PBH DM, decaying, and annihilating particle DM we perform the above analysis and obtain the corresponding sensitivities. 

In Fig.\,\ref{fig:-PBH}, we show the sensitivities on evaporating PBH DM with MeVCube of different sizes. Here the PBHs are assumed to be non-spinning and of monochromatic mass distribution. Other existing limits in the parameter space are taken from Refs.\,\cite{Laha:2019ssq, DelaTorreLuque:2024qms, PhysRevLett.126.171101,Laha:2020ivk, Dasgupta:2019cae,PhysRevD.101.023010, Chen:2021ngo, Laha:2020vhg,Kim:2020ngi,Clark:2016nst} (see figure caption for more details). Besides these, some other relevant limits and sensitivities in the parameter space are: Super-Kamiokande limit on diffuse supernovae neutrino background (DSNB)\,\cite{Dasgupta:2019cae,Bernal:2022swt}, sensitivities from upcoming telescopes like JUNO\,\cite{Wang:2020uvi,Liu:2023cqs}, DUNE\,\cite{DeRomeri:2021xgy}, global 21 cm measurement\,\cite{Clark:2018ghm, Mittal:2021egv, Saha:2021pqf, Natwariya:2021xki}, Lyman-$\alpha$ forest measurements\,\cite{Saha:2024ies}, LeoT dwarf gas heating\,\cite{Laha:2020vhg, Kim:2020ngi}, cosmic ray measurements from Voyager 1\,\cite{DelaTorreLuque:2024qms, Boudaud:2018hqb}, and radio measurements from the inner GC region\,\cite{Chan:2020zry}. The sensitivities from MeVCube go beyond the current limits in the PBH DM parameter space. In addition, with increasing size of MeVCube, the effective area increases and as a result the reach of the telescopes also improves as can be seen in Fig.\,\ref{fig:-PBH}. Note that the limits on PBH DM extend below $5\times10^{16}$ g. We do not show them here because at those lower masses the 511 keV limits are much stronger\,\cite{DelaTorreLuque:2024qms}. We note that PBHs in the mass range $10^{18}$ g to $10^{22}$ g can make up large fraction of the total DM density. Some ideas for probing this mass window can be found in  \cite{1992ApJ...386L...5G,Katz:2018zrn,Nemiroff:1995ak,Jung:2019fcs,Bai:2018bej,Laha:2018zav,Montero-Camacho:2019jte,Ghosh:2022okj,Tran:2023jci,Gawade:2023gmt,Tamta:2024pow,Crescimbeni:2024cwh,Thoss:2024dkg,Crescimbeni:2024qrq}. 

The results for particle DM decaying and annihilating to two photons, are shown in Fig.\,\ref{fig: -FSR decay and annihilation} (top row). The previous limits are taken from Refs.\,\cite{Laha:2020ivk,Siegert:2021upf,Essig:2013goa,Slatyer:2015jla} (see figure caption for more details). We see that MeVCube has the potential to probe longer lifetimes (for DM decay) and smaller cross-sections (for DM annihilation), compared to all the existing limits for both decaying and annihilating DM. In Ref.\,\cite{Lucchetta:2022nrm} the effective area of a 12U MeVCube is shown to extend to higher energy photons, beyond 2U and 6U. As a result, we see in Fig.\,\ref{fig: -FSR decay and annihilation}, with a 12U MeVCube we are able to probe heavier masses of particle DM. The same effect is also present for PBH DM in Fig.\,\ref{fig:-PBH}, where a 12U MeVCube will be sensitive to lower masses of PBH DM (the peak energy of emitted photons increases with decreasing PBH mass). In Fig.\,\ref{fig:-PBH} we do not show this advantage of 12U MeVCube, as the parameter space for low-mass PBH DM is already strongly ruled out. 

In Fig.\,\ref{fig: -FSR decay and annihilation} 
 (bottom row) we show the limits for particle DM decaying and annihilating to electron-positrons which in turn produce photons due to FSR. The previous limits are obtained from Refs.\,\cite{Liu:2016cnk,Calore:2022pks,Essig:2013goa,Wadekar:2021qae,Boudaud:2016mos,Cirelli:2023tnx,Slatyer:2015jla,DelaTorreLuque:2023olp,DelaTorreLuque:2023cef} (see figure caption for more details). Given the DM is not directly decaying or annihilating to photons, the flux of photons expected in this case is much lower compared to direct DM decay or annihilation to photon final states. As a result, sensitivities from MeVCube are weaker compared to the existing limits and in future can be used as a complementary probe for the electron-positron channel of DM decay and annihilation. We note that FSR limits in Fig.\,\ref{fig: -FSR decay and annihilation} can be extended to higher DM masses. Given the stronger existing limits in the parameter space, we choose not to extend the plot.

To cross-check our results, we have tried to reproduce the results in Ref.\,\cite{Ray:2021mxu}. We were able to reproduce their PBH DM sensitivities for AMEGO-X. We note that the limits presented in this work are competitive with the sensitivity studies for other proposed MeV telescopes\,\cite{Coogan:2020tuf, Coogan:2021sjs, Caputo:2022dkz, ODonnell:2024aaw}. Many of the previous sensitivity studies use different forecast analysis, sky region, and observation time across all different telescopes. We leave a more robust comparison between all different telescopes for future analysis.

 \section{Discussion and scope}
\label{scope}

In this work we have shown the capability of the proposed MeVCube satellite, in the search for DM. We consider low-mass evaporating PBH DM and decaying, annihilating particle DM as possible DM candidates and calculate the resulting photon flux from them. Using the effective area and energy resolution of MeVCube, we evaluate the sensitivities of MeVCube for these DM models. In almost all cases, we find that MeVCube will be able to probe new regions of parameter space that are beyond the reach of current telescopes. 

For PBH DM, we have considered non-spinning PBHs with monochromatic mass distribution. We leave the exploration of spinning PBH with other mass distributions for future work. For particle DM decaying or annihilating to electron-positrons, galactic propagation effects on the charged particles are important for evaluating the resulting secondary photon flux. Besides heavier particle DM particles can directly decay or annihilation to electron-positrons which then can upscatter low-energy background photons and produce photons in the energy window of MeVCube. Positrons produced from DM decay, annihilation, and PBH DM evaporation, can also contribute to the galactic 511 keV excess\,\cite{DelaTorreLuque:2024wfz}. Given the uncertainty in the various propagation models in the literature, a detailed future study will be interesting.

In our work, we have obtained the sensitivities assuming the galactic and extragalactic diffuse background. For MeVCube, which is planned to be launched at the low Earth orbit (LEO), there are other backgrounds like cosmic ray charged particles, albedo photons, and instrumental backgrounds\,\cite{Cumani:2019ryv} that one must take into account. Thus with a detailed background modeling of
MeVCube, one can evaluate a more robust limit for different DM models. 

Besides GC, some other interesting targets for DM search include dwarf galaxies, M31 galaxy, and galaxy clusters. Several works\,\cite{Coogan:2020tuf, Coogan:2021rez,Caputo:2022dkz, ODonnell:2024aaw} have shown that for the proposed MeV telescopes, GC limits are stronger than the limits from other sources, given the larger signal-to-noise ratio for GC. The study of MeVCube's DM sensitivities for other DM-rich sources can provide important complementarity. 
 
MeVCube can also be sensitive to strongly interacting DM. Space-based missions have been previously used to search for such DM candidates\,\cite{Bhoonah:2020fys,Li:2022idr,Wandelt:2000ad, Alpine:2024kej, Du:2024afd}. A detailed analysis for strongly interacting DM will be insteresting for MeVCube and other upcoming MeV telescopes.
 
 The MeVCube collaboration has not given any timeline yet, though according to Ref.\,\cite{Lucchetta:2022nrm}, a prototype has already been tested at DESY. One can still compare the development timeline of other CubeSat telescopes with big scale satellites. For example, NinjaSat is a 6U CubeSat X-ray telescope launched in 2023. The timeline of this launch can be seen in Fig.\ 12 of Ref.\,\cite{Tamagawa:2024kze}. The whole project, from development to launch, was completed within 3 years. In comparison, Ref.\,\cite{cosiTime} notes that starting from 2021, COSI is still under development. COSI is scheduled to be launched in 2027. MeVCube has size comparable to NinjaSat, and thus it is also expected to be launched faster than big-scale missions.

 Already there are many existing and proposed CubeSat missions\,\cite{Brown2012TheCX, article, Mason:2015mna,Johnson:2015tga,Chattopadhyay:2018yek,s17030493,Kaaret:2019xvm,Yang:2019bqd, Feng:2019hma, Yatsu2019CubeSatFU, Fuschino:2018ddv,Elsaesser2020SpectroCubeAE,fabiani2024cubesat,Solomey:2022gja,france2023colorado, Bloser:2022pnu, Braga:2020vnv, wen2021compact, 10.1117/12.2595859,Kushwah:2021qnb, ardila2022star, LEHTOLAINEN2022166865, knapp2020demonstrating, raskin2018cubespec,Pal:2023qjo, CAMELOT:2018bqg,Zhu:2023zwr,Diwan:2023imi,deKuijper:2024qgv,Tamagawa:2024kze,Lacour:2018rnl} for photon, neutrino and gravitational wave astronomy. Investigation of New Physics capabilities of these missions can be an important avenue. With the advent of new detector technologies and understanding of the various astrophysical backgrounds, we hope to uncover the non-gravitational nature of DM soon.

\section{Acknowledgements}
I especially thank Ranjan Laha and Nirmal Raj for extensive discussions and helpful comments on the manuscript. I also thank Biplob Bhattacherjee, Bhavesh Chauhan, Marco Cirelli, Andrea Caputo, Amol Dighe, Raghuveer Garani, Joachim Kopp, Jason Kumar, Sudhir Kumar Vempati, Alejandro Ibarra, Rinchen Sherpa, Tracy Slatyer, Rahul Srivastava, Ujjwal Upadhyay, Anupam Ray, Robin Anthony-Petersen, and Yu Watanabe for useful discussions. I acknowledge
the Ministry of Human Resource Development, Government of India, for financial support via the Prime
Ministers’ Research Fellowship (PMRF). 
\appendix 
\section{Fisher Forecasting analysis}
\label{appendix}
One of the most popular ways of estimating DM sensitivities for an upcoming telescope is to simply look at the signal-to-background ratio. If $N_\gamma^{\rm DM}$ is the number of photon events detected in some time interval and the corresponding background event number is $N_\gamma^{\rm bkg}$, then for potential discovery one requires,
\begin{eqnarray}
    \frac{N_\gamma^{\rm DM}}{\sqrt{N_\gamma^{\rm bkg}}} \geq n_s\,\,,
\end{eqnarray}
where $n_s$ is the significance with which one wants to make the discovery. For example, in case of $5\sigma$ discovery, $n_s=5$. 

One important limitation of this sort of approach is that it assumes a constant uncertainty value for the fiducial background parameters and just searches for a DM signal over background. For example, if we look at the background model in Eq.\,(\ref{galbkg}), the fiducial value and the error-bars are obtained by fitting with COMPTEL telescope results. After the launch of MeVCube, the uncertainties in the background parameter values may change because MeVCube will make an independent measurement of the galactic background flux. Thus for estimation of New Physics capabilities of upcoming telescopes, one also has to take into account the changes in both the model and background parameters. For these reasons, Ref.\,\cite{Edwards:2017mnf} introduced Fisher forecast analysis for DM search analysis with upcoming telescopes.

Fisher forecasting is one of the most popular tools used for sensitivity estimates in cosmology and astrophysics\,\cite{Albrecht:2006um,Wittman,2009arXiv0901.0721A,2009arXiv0906.4123C,Wolz:2012sr,Lamperstorfer:2015cfg}.
The method relies on the estimation of Fisher information matrix. Given the exposure of a telescope one can then determine how constraining a future observation will be for all the model and background parameters involved in the problem. 

The Fisher matrix provides error estimates of the parameters as well as the covariances between them. These quantities can be calculated from the Fisher matrix itself. The correlation coefficient is given by,
\begin{eqnarray}
    \rho_{ij}=\frac{C_{ij}}{\sqrt{C_{ii}\times C_{jj}}}\,\,,
    \label{correlation}
\end{eqnarray}
where $C_{ij}$ is the covariance matrix element related to the Fisher matrix element via
\begin{eqnarray}
    C_{ij}= \mathcal{F}_{ij}^{-1}.
\end{eqnarray}

The correlation coefficient lies within $[-1,1]$, where  $\rho_{ij}>0$ means positive correlation, $\rho_{ij}<0$ means negative correlation, and $\rho_{ij}=0$ is zero correlation. The correlation coefficient helps in understanding the degeneracies in the parameters involved. For example, if two parameters $\mathcal{A}$ and $\mathcal{B}$ are negatively correlated, then that implies that a decrease in parameter $\mathcal{A}$ can be compensated by an increase in parameter $\mathcal{B}$ to get the same fit to the observation. This makes the independent determination of parameters $\mathcal{A}$ and $\mathcal{B}$ difficult.

\begin{figure}[htbp]
    \centering
    \includegraphics[width=\textwidth]{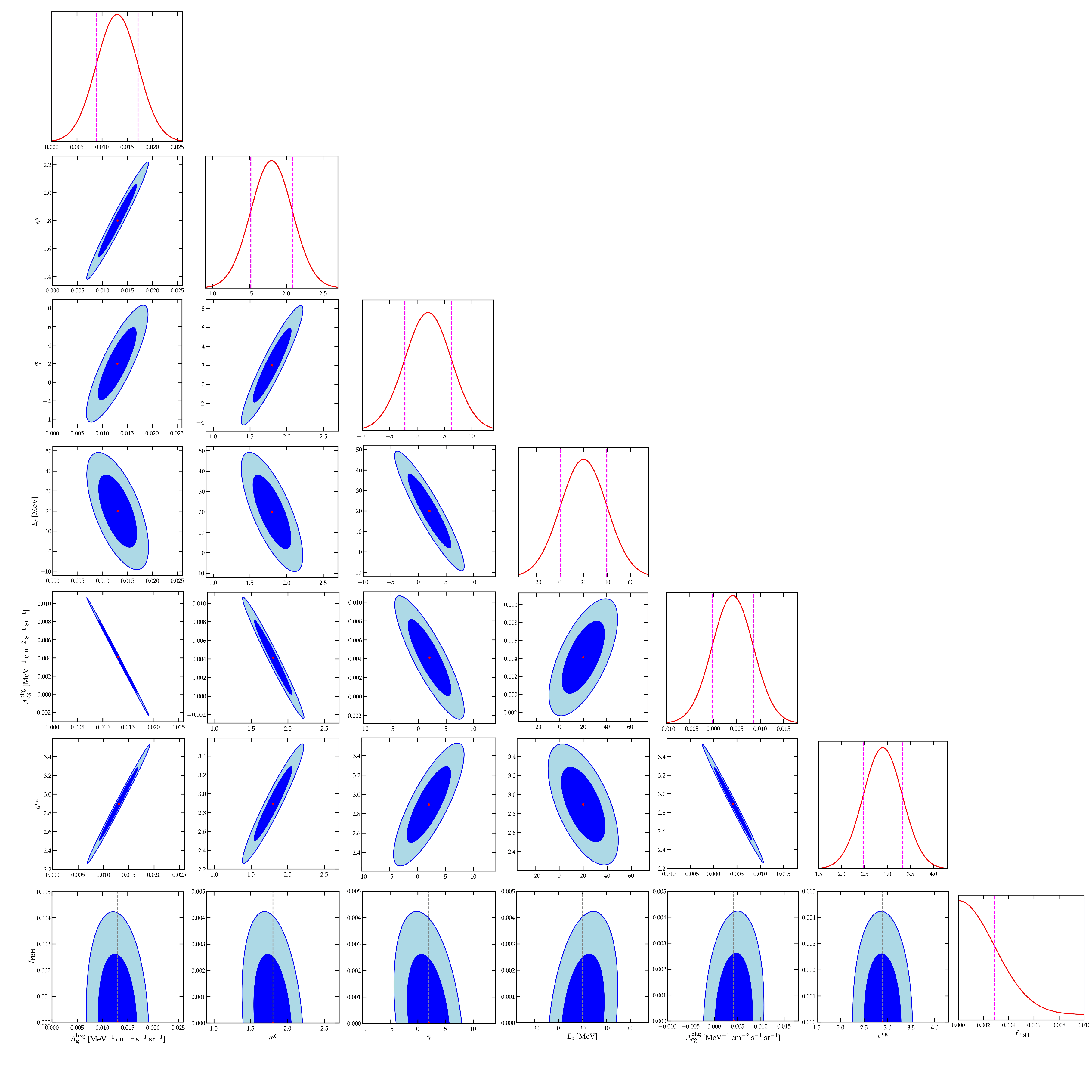}
    \begin{minipage}{\textwidth} 
        \caption{Fisher forecast constraints on the model and the background parameters. This plot assumes a non-spinning PBH DM of mass $10^{17}$g and an observation time of 10$^7$ s with a 12U MeVCube. The central point of each plot represents the mean (fiducial) values of the parameters involved and the vertical violet dashed lines show the 95\% uncertainty bound. The corresponding 95\% (68.3\%) CL confidence ellipses are shown by the light (deep) blue curves. The vertical grey dashed line in the bottom row show the zero correlation coefficient line. For the minimum variance on $f_{\rm PBH}$ (bottom row, last panel), we only show the one sided distribution as negative $f_{\rm PBH}$ values are unrealistic.}
        \label{fig:-PBHFisher}
    \end{minipage}
\end{figure}

\begin{figure}[htbp]
    \centering
    \includegraphics[width=\textwidth]{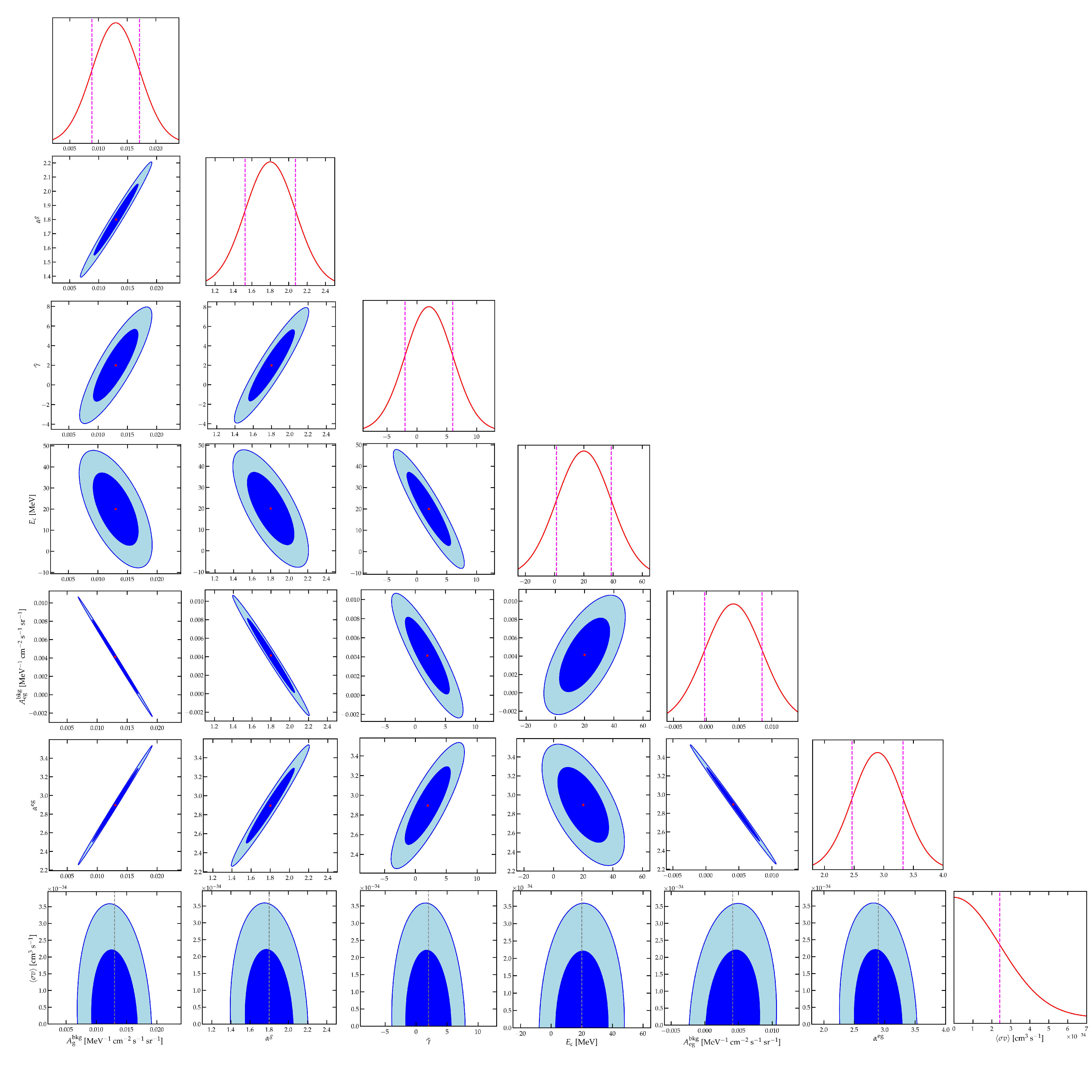}
    \begin{minipage}{\textwidth} 
        \caption{Same as the previous figure, but now for particle DM of mass 1 MeV annihilating directly to two photons.}
        \label{fig:-AnnihilationFisher}
    \end{minipage}
\end{figure}

In Fig.\,\ref{fig:-PBHFisher} we show the corner plots with individual uncertainties and covariances for the signal and background model parameters. For this plot we have chosen a 12U MeVCube and $t_{\rm obs}=10^7$ s. Here the signal is the photon flux from evaporating PBH DM of mass $10^{17}$ g and the background model parameters are those described in Eqs.\,(\ref{galbkg},\,\ref{exgalbkg}). For example, the first column in Fig.\,\ref{fig:-PBHFisher} shows (from top to bottom) the parameter uncertainty in $A_{\rm g}^{\rm bkg}$ and the covariances of $A_{\rm g}^{\rm bkg}$ with $\alpha^{\rm g},\bar{\gamma}, E_c,  A_{\rm eg}^{\rm bkg}, \alpha^{\rm eg},$ and $f_{\rm PBH}$, respectively. The correlation coefficients for DM parameter $f_{\rm PBH}$ with the background parameters are calculated using Eq.\,(\ref{correlation}). The correlation coefficient values are: $-0.146$ ($f_{\rm PBH}-A_{\rm g}^{\rm bkg}$), $-0.269$ ($f_{\rm PBH}-\alpha^{\rm g}$), $-0.347$ ($f_{\rm PBH}-\bar{\gamma}$), $0.306$ ($f_{\rm PBH}-E_c$), $0.137$ ($f_{\rm PBH}-A_{\rm eg}^{\rm bkg}$), and $-0.055$ ($f_{\rm PBH}-\alpha^{\rm eg}$). For a representative of a line photon signal, we also show in Fig.\,\ref{fig:-AnnihilationFisher} the corner plots for 1 MeV particle DM annihilating to two photons. 
\section{Dependence on DM density profile and detection time}
Besides the widely used NFW profile there are various other DM density profiles in the literature. Here we consider the Isothermal and the Einasto profile.  For Isothermal profile Eq.\,(\ref{NFW}) admits the following values, $(a, b, c) =(2, 2, 0)$ and $r_s = 3.5$ kpc.

\begin{figure}
	\begin{center}
	\includegraphics[height=7.8cm]{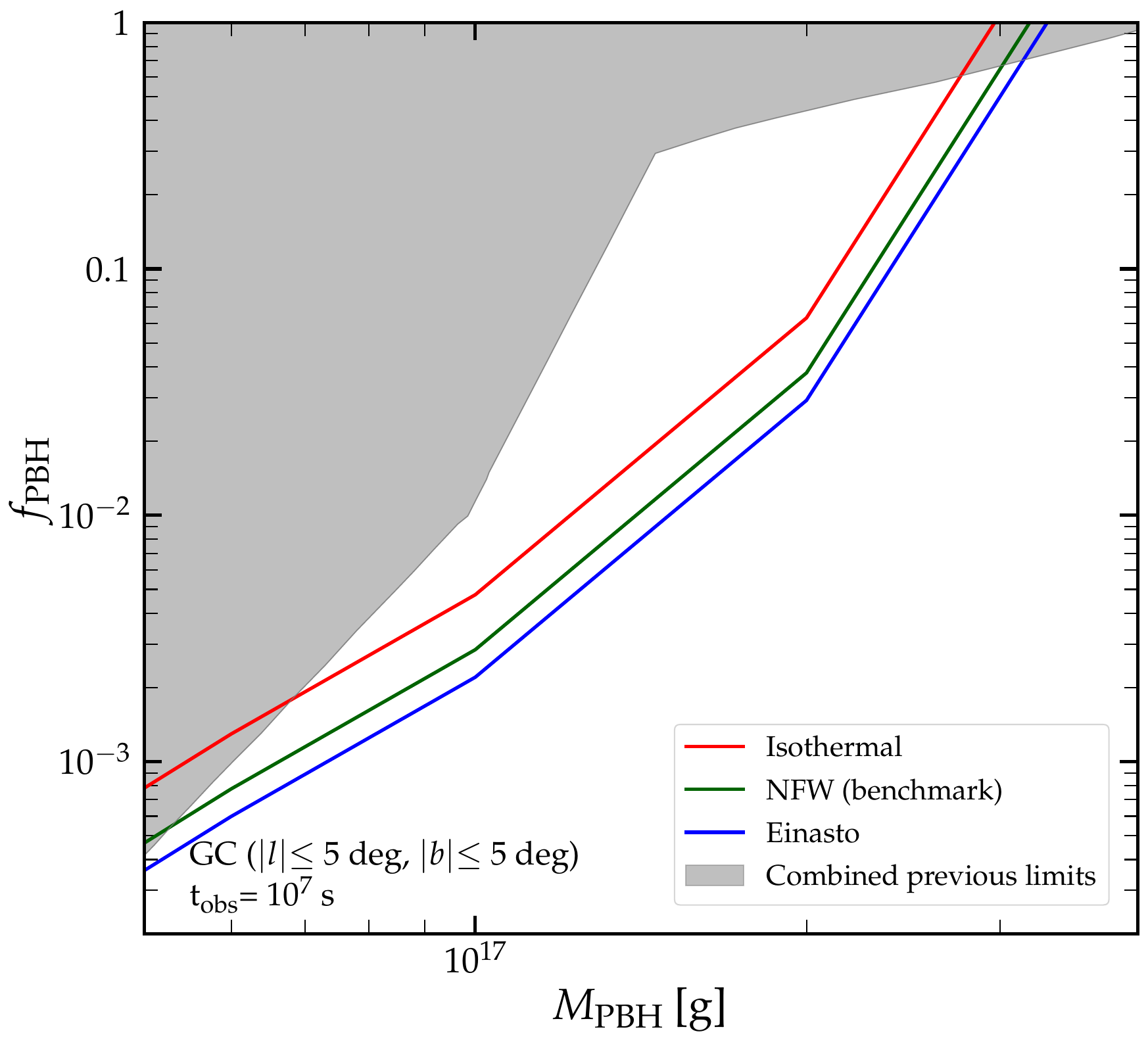}
	\caption{Dependence of PBH DM sensitivities with a 12U MeVCube (observation time $10^{7}$ s ) on various DM density profiles. Limits for the isothermal, NFW (benchmark), and Einasto profiles are shown by red, green, and blue solid curves, respectively. The combined previous limit is shown by the grey shaded region.}
	\label{fig:-LimitProfile} 
     \end{center}
\end{figure}

Einasto profile is given by,
\begin{eqnarray}
    \rho_\chi^{Ein} (r) \,=\, \rho_\odot \, {\rm exp}\left(-\frac{2}{0.17}\frac{r^{0.17}-(8.3 \,\rm kpc)^{0.17}}{(20 \,\rm kpc)^{0.17}}\right)\,.
\label{Einasto}
\end{eqnarray}
We evaluate the sensitivities on low-mass evaporating PBH DM from a 12U MeVCube for Isothermal and Einasto profiles in Fig.\,\ref{fig:-LimitProfile}. For comparison we also show the benchmark NFW density profile used in this work. Isothermal profile predits a `cored' profile towards GC as opposed to the `cuspy' profiles like NFW or Einasto. These relative density profiles explain the strengths of the limits in Fig.\,\ref{fig:-LimitProfile}, where Einasto profile provides the strongest limits whereas Isothermal provides the weakest limit. 

\begin{figure}
	\begin{center}
	\includegraphics[height=7.8cm]{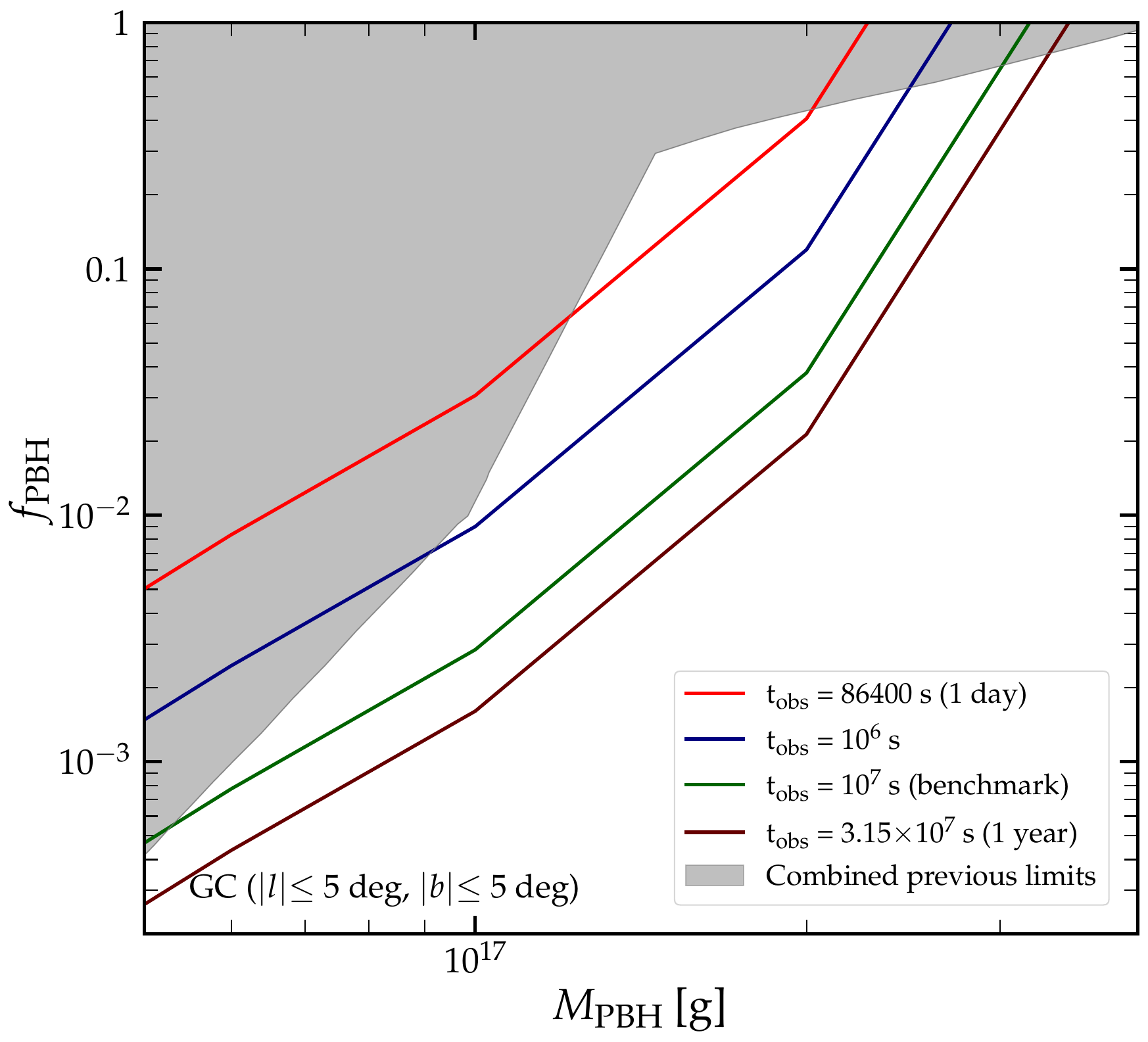}
	\caption{Dependence of PBH DM sensitivities on various observation times with a 12U MeVCube (NFW density profile). Limits for observation times of 1 day, $10^6$ s, $10^7$ s (benchmark), 1 year are shown by red, blue, green, and brown solid lines, respectively. The combined previous limit is shown by the grey shaded region.}
	\label{fig:-LimitLifetime} 
     \end{center}
\end{figure}

Another important consideration for obtaining limits on DM is the time of observation. As a fiducial value we have chosen $t_{\rm obs}= 10^7$ s, which corresponds to around 3 months and 26 days of observation. We choose this value in accordance with Ref.\,\cite{Lucchetta:2022nrm}, which estimates $\sim$ 2
months observation time at a given source, for MeVCube. Given the short lifetimes of many of the CubeSat missions one can ask the question: can we learn anything important about DM with a shorter telescope observation time? In Fig.\,\ref{fig:-LimitLifetime} we show the dependence of our results on observation time of 
 a 12U MeVCube for the case of evaporating PBH DM. The limits here scale as $\sqrt{t_{\rm obs}}$. We see that even if MeVCube observes the GC region of interest considered here ($|l| < 5^{\circ}, |b| < 5^{\circ}$) for 1 day, one will be able to explore new regions of PBH DM parameter. Similarly, for longer observation times, the limits will improve accordingly. The same conclusions should be true for both particle DM decay and annihilation.  
\section{Regarding the cut-off energy in the galactic background model}

For galactic photon background model we used the analytical form provided in Ref.\,\cite{Bartels:2017dpb}, 
 \begin{eqnarray}
     \phi_{\rm g}^{\rm bkg}(E_\gamma)= A_{\rm g}^{\rm bkg}\left(\frac{E_\gamma}{1\,\rm MeV}\right)^{-\alpha^{\rm g}}{\rm exp}\left(-\left(\frac{E_\gamma}{E_c}\right)^{\bar{\gamma}}\right)\,\,,
     \label{Appgalbkg}
 \end{eqnarray}

 where the fiducial values of all the parameters above are given below Eq.\,(\ref{galbkg}). The exponential cut-off energy in the above equation is, $E_c=20$ MeV. 

\begin{figure}
	\begin{center}
	\includegraphics[height=7.8cm]{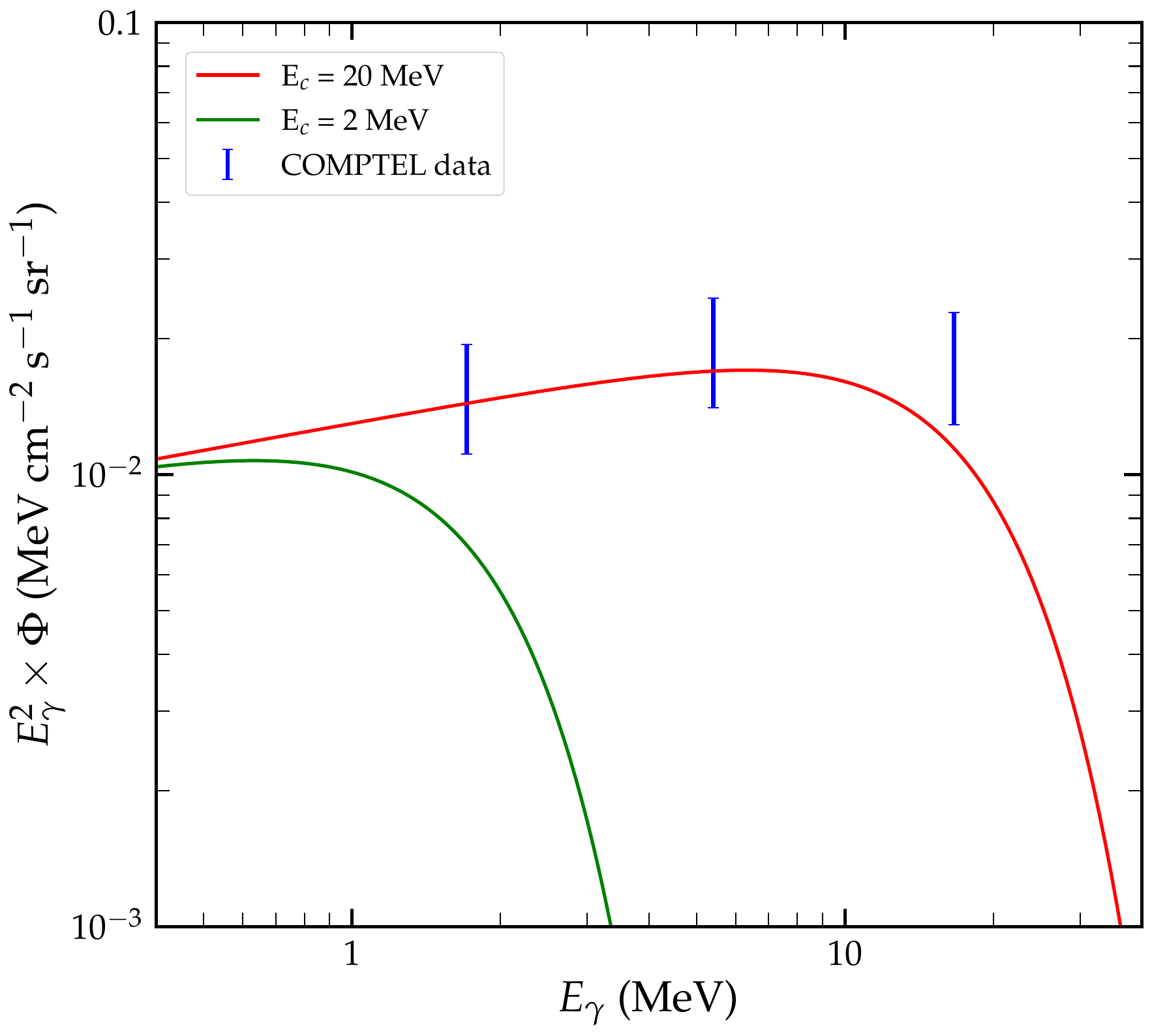}
	\caption{Comparison of the analytical galactic background model used in this work with the COMPTEL data from the inner galaxy ($|l| < 30^{\circ}, |b| < 5^{\circ}$). The background models are shown for $E_c=20$ MeV and $E_c=2$ MeV by solid red and solid green lines, respectively. The background model is in agreement with the COMPTEL data for $E_c=20$ MeV.}
	\label{fig:-Cutoff} 
     \end{center}
\end{figure}

 During this work we found a typo in Ref.\,\cite{Bartels:2017dpb}, more specifically in Eq.\,(4.1) of the paper. From that equation it seems that the cut-off energy for the galactic background is 2 MeV. This is not compatible with Fig.\,(2) of their paper (`ICS$_{\rm lo}$' component). In Fig.\,\ref{fig:-Cutoff} we compare the galactic background model with the COMPTEL measurements from the inner part of the galaxy, $|l| < 30^{\circ}, |b| < 5^{\circ}$\,(Fig.\,8 of \cite{2005A&A...444..495S}) for $E_c= 20$ MeV and $E_c= 2$ MeV. Clearly the data is in agreement with the model having $E_c= 20$ MeV cut-off and not with $E_c= 2$ MeV. 

 We note that this typo is also present in the analysis of Ref.\,\cite{Coogan:2021rez}. The limits on PBH DM, for the parameter space of interest, is not very sensitive to this typo of $E_c$. But for particle DM decay and annihilation, the assumption of $E_c=2$ MeV overestimates the resulting limits at heavier masses of particle DM. We emphasize this typo here so that there is no confusion in any future analysis.  

\section{Launch cost and frequency for MeVCube}

The timeline for MeVCube is yet to be public. So the launch vehicle and launch dates are uncertain. But given the rise of many independent organizations for CubeSat launch, one can do a rough estimate of how cost-effective MeVCube will be compared to other large-scale telescopes. Here we use the launch cost estimates provided for the Space-X `rideshare program' (\href{https://www.spacex.com/rideshare/}{https://www.spacex.com/rideshare/}). For a launch at low-earth orbit, the approximate cost is 0.33 million dollars for CubeSat payload below 50 kg. Given a 1U MeVCube weights around 1.33 kg\,\cite{Lucchetta:2022nrm}, we show the weights and corresponding launch costs for 2U, 6U, and 12U MeVCube in table \ref{tab:launch cost}. For all the three CubeSat sizes, the launch cost is 0.33 million dollars (given this is the minimum cost for launch with Space-X). In contrast, for COSI the launch cost is around 69 million dollars as stated in Ref.\,\cite{cosiCost}.

\begin{table}[h!]
\begin{tabular}{ |p{7em}|p{7em}|p{7em}| } 
\hline
\vspace{0.2mm} CubeSat units & \vspace{0.2mm} Weight (in kg) & \vspace{0.2mm} Launch cost (in US dollars)   \\
\hline 
\vspace{0.2mm}
2U & \vspace{0.2mm} 2.66  & \vspace{0.2mm} 0.33 million \\ 
\hline
 \vspace{0.2mm}6U & \vspace{0.2mm} 7.98  &  \vspace{0.2mm} 0.33 million \\ 
\hline
\vspace{0.2mm} 12U &  \vspace{0.2mm} 15.96  &  \vspace{0.2mm} 0.33 million\\ 
\hline
\end{tabular}
\caption{Estimate of weight and corresponding launch cost for different sizes of CubeSat, obtained from \href{https://www.spacex.com/rideshare/}{https://www.spacex.com/rideshare/}}

\label{tab:launch cost} 
\end{table}

 MeVCube suffers from small effective area as compared to big-scale telescope proposals like AMEGO-X. Besides, many of the launched CubeSat satellites have smaller lifetimes. Given these limitations, one can estimate how many CubeSats like MeVCube are required to achieve the same sensitivities as AMEGO-X. The effective area for a 12U MeVCube at 1 MeV photon energy is $\sim$ 22 cm$^2$. At the same energy the effective area for the proposed AMEGO-X is $\sim$ 3325 cm$^2$. Thus to achieve the same effective area as AMEGO-X at 1 MeV $\sim$ 151 such CubeSats are required. Given the above cost estimates, launching 151 such 12U CubeSats will cost a total of $\sim$ 50 million dollars. According to Ref.\,\cite{AMEGO:2019gny} (table 2), the launch cost for AMEGO-X is estimated to be around 100 million dollars. So just from the launch cost perspective the CubeSat option seems more cost-effective. A more realistic comparison would be very interesting for the science case of MeVCube.

Different CubeSats are launched multiple times in a year. According to Ref.\,\cite{numCubesat} around 4400 CubeSats have been launched till date. The frequency of these launches can be seen in Ref.\,\cite{CubeFrequency}. There are many independent organizations like SpaceX\,\cite{SpaceX}, RocketLab\,\cite{RocketLab} that do CubeSat launches frequently. Therefore MeVCube, if approved, can be launched much quicker than large-scale telescopes.
 
\section{Choice of orbit for MeVCube}

Historically, low-earth orbit (LEO) (altitudes $\sim$ 550 km) has been chosen for most of the gamma-ray satellites. One reason for that is the lower power required to put satellites in LEO compared to farther orbits (like high-earth orbit). Cosmic ray charged particles can increase the background count of any gamma-ray telescope and even damage the instruments. For LEO, the earth's magnetic field can shield the satellites from cosmic rays. An estimate of the background flux of proton, alpha-particles and electron-positrons at LEO compared to the interplanetary medium is shown in Fig.\,9 of Ref.\,\cite{Tatischeff:2022mwc}. In the presence of the geomagnetic field, we see that below 1 GeV/nucleon energy the cosmic ray flux is highly suppressed in LEO compared to near-Earth interplanetary medium. All future MeV telescopes including MeVCube will operate below this energy. Even for LEO, the cosmic ray flux increases with altitude, as can be seen in Fig.\,3 and 4 of Ref.\,\cite{vripa2020comparison}. Thus putting a satellite in farther orbits than LEO can affect its sensitivity by increasing the background. However, we note that LEO passes through Van Allen radiation belts and the South Atlantic Anomaly (SAA), where the background radiation count can significantly increase. For long, uninterrupted observations, some telescopes thus choose father orbits even though the overall background flux there is greater than LEO. One such example is the INTEGRAL satellite which was placed at the highly elliptical orbit (HEO)\,\cite{IntegralOrbit}. Thus the choice of orbits for MeV satellites depends on the planned duration of the mission and background mitigation strategies.

\clearpage
\clearpage

\clearpage
\clearpage
\bibliographystyle{JHEP}
\bibliography{ref.bib}

\end{document}